# High sensitivity silicon carbide divacancy-based thermometer

Qin-Yue Luo[1], Shuang Zhao[1], Qi-Cheng Hu[1], Wei-Ke Quan[1], Zi-Qi Zhu[1], Jia-Jun Li[1], Jun-Feng Wang[1, *]

[1]*College of Physics, Sichuan University, Chengdu, Sichuan 610065, People's Republic of China*

*Corresponding author: jfwang@scu.edu.cn

**Abstract**

**Color centers in silicon carbide have become potentially versatile quantum sensors. Particularly, wide temperature range temperature sensing has been realized in recent years. However, the sensitivity is limited due to the short dephasing time $T_2^*$ of the color centers. In this work, we realize a high sensitivity silicon carbide divacancy-based thermometer using the thermal Carr-Purcell-Meiboom-Gill (TCPMG) method. First, the zero-field splitting D of PL6 divacancy as a function of temperature is measured with a linear slope of -99.7 kHz/K. The coherence times of TCPMG pulses linearly increase with the pulse number and the longest coherence time is about 21 μs, which is ten times larger than $T_2^*$. The corresponding temperature sensing sensitivity is 13.4 mK/Hz$^{1/2}$, which is about 15 times higher than previous results. Finally, we monitor the laboratory temperature variations for 24 hours using the TCMPG pulse. The experiments pave the way for the applications of silicon carbide-based high sensitivity thermometer in the semiconductor industry, biology, and materials sciences.**

High sensitivity nanoscale thermometer is significant for various fields of science and technologies [1,2]. Multiple traditional methods including scanning thermal microscopy, Raman spectroscopy and fluorescent protein thermometry have been widely applied to microscale temperature detection in biological science, materials and electronic devices [1,2]. However, these methods suffer from low sensitivity or unstable fluorescence. In the past decades, color centers in diamond have been developed to high sensitivity nanoscale quantum thermometers [1,2]. The principles of the color centers-based thermometers are temperature dependent zero-field-splitting (ZFS) and fluorescence spectra [2]. They exhibit many advantages such as high sensitivity [2], wide temperature detection range [3], ultra-stable fluorescence, and biocompatibility [1-3]. Recently, the temperature sensing sensitivity has been improved to about 100 $\mu K/Hz^{1/2}$ using the hybrid diamond nanothermometer [4]. The diamond thermometer has a range of temperature detection applications in biological science such as living cells and thermogenetic neurostimulation, nanoscale thermal conductivity imaging, and electronics including semiconductor devices and coplanar waveguide thermal imaging [2].

However, both the excitation laser and the fluorescence of color centers in diamond are in the visible range, which will cause larger auto-fluorescence and optical absorption of living cells [5]. In addition, for temperature measurements in electronics, diamonds should be coated on the electronic devices, which is difficult in situ to detect the temperature of the electronics devices [2,6]. The diamond is expensive and hard to fabricate, which limits the wide applications of diamond thermometers [7-9]. As a beneficial complement, color centers in silicon carbide (SiC) have been used as thermometers, which have near-infrared fluorescence and long coherence time even at room temperature [10-12]. SiC is a semiconductor with extensive applications in high-power microelectronic devices, which makes it possible for the intrinsic color centers in-situ quantum sensing including the temperature detection [7-11]. It has mature inch-scale growth and fabrication technologies, which is convenient for the application of the SiC-based quantum sensing.

In recent years, various color centers have been observed in SiC, which could be divided into two types, bright single photon emitters [13] and spin qubits [7-9,14,15]. There are three types of spin qubits containing divacancy [7,8,10,11], silicon vacancy [9,12] and NV centers [14,15], whose spin states can be polarized and controlled by laser and microwave, respectively [7-9]. The spin qubits have been used in quantum photonics [16,17], quantum computation [18,19], and quantum sensing. It can be developed to sense electric field [20], magnetic field [21], strain [22], temperature [10-

12] and high pressure [23,24]. Ramsey method has been used for wide-temperature range temperature sensing, but the sensitivity is limited to about 200 mK/Hz$^{1/2}$ due to the short dephasing time T$_2^*$ [11,12]. In order to improve the sensitivity, the dynamical decoupling methods need to be used to increase the coherence time [25,26].

In this work, we realize SiC divacancy high sensitivity thermometer using the thermal Carr-Purcell-Meiboom-Gill (TCPMG) method at room temperature. First, the temperature dependent divacancy ZFS is measured, showing that it linearly decreases with temperature at a slope of -99.7 ± 0.4 kHz/K. On this basis, we measured the TCPMG under different pulse numbers N, and the coherence times increase with the pulse number. The maximum sensing sensitivity reaches 13.4 mK/Hz$^{1/2}$, which is about 15 times higher than previous results. Finally, we use the TCPMG methods to monitor the laboratory temperature for one day. The experiments demonstrate a high sensitivity silicon carbide divacancy-based thermometer, which can be applied to biology, materials science and semiconductor.

There are seven types of divacancies in 4H-SiC: PL1- PL7 [7,8], and they exhibit spin-1 ground spin state. They can be divided into two types with two directions: one is c-axis, including PL1, PL2, and PL6, and the other is basal, including PL3-PL5, and PL7 [7,8]. Since the TCPMG pulse sequences need simultaneously controlling the transitions of $|0\rangle \leftrightarrow |1\rangle$ and $|0\rangle \leftrightarrow |-1\rangle$ at a magnetic field [25,26], we select the PL6 divacancy in the experiments. The spin Hamiltonian of the PL6 divacancy defects is [25-27]

$$H = DS_Z^2 + g\mu_B \mathbf{B}\cdot\mathbf{S} + \mathbf{S}H_B, \qquad (1)$$

where *D is the* zero-field-splitting (ZFS) parameter, which is temperature dependent. The D value is about 1350 MHz at room temperature [7, 21]. *g = 2* is the electron g-factor, $\mu_B$ denotes the Bohr magneton, ***B*** denotes the applied axial static magnetic field, and the spin **S** = 1. $H_B$ denotes the interaction between divacancy and spin bath.

In our study, a confocal system combined with microwave and magnetic systems is utilized [10,11,21]. The sample is a high-purity semi-insulating 4H-SiC (Cree) with ensemble divacancy [7,10,11]. An acousto-optic modulator (AOM) is used to modulate the 914 nm laser into laser pulses. After being reflected by a 980 nm dichroic mirror, the laser focuses on the sample through an objective. The fluorescence is collected by the same objective, then it is filtered by a 1000 nm long-pass filter. The multimode fiber is utilized to couple the light to a photoreceiver (Femto, OE-200-IN1). Two microwave generators controlled by a pulse generator (PBESR-PRO-500, Spincore) are used to manipulate the two transitions. To transfer microwaves and control the spins of divacancy, a copper wire with a 50 μm diameter is placed on the sample. The lock-in methods are performed to detect the spin signals [7,10,14]. We apply a metal ceramic heater (HT24S, Thorlabs) and a platinum resistive temperature sensor (TH100PT, Thorlabs) to control and investigate the sample's temperature, respectively. An electromagnet is kept underneath the sample throughout the experiments to produce a coaxial magnetic field.

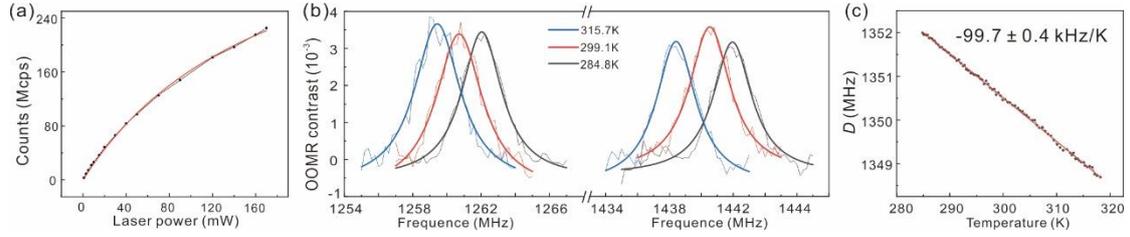

**Figure 1.** ODMR measurements. **(a)** Saturation curve. The red line is a fitting curve to the data. **(b)** ODMR measurements at three representative temperatures at a magnetic field of 32.1 G, respectively. The blue, red, and gray lines are Lorentz fits at 315.7 K, 299.1 K, and 284.8 K, respectively. **(c)** The ZFS D as a function of the temperature from 280 K to 320 K. The red line is a linear fitting to the data, which demonstrates that the ZFS D has a linear dependence on temperature.

Since the fluorescence count is an important parameter determining the sensing sensitivity, we first examine the saturation curve of the divacancy ensemble. As displayed in Fig. 1a, the counts rise together with the laser power. A function $I(P)=I_s/(P_0/P+1)$ is used to fit the data (Red line), where $I_s$ is the saturation count of the sample and $P_0$ is the saturation laser power. From the fit, we get that the $I_s$ = 458 Mcps and $P_0$ = 182 mW. The measured maximum count is 224 Mcps at 170 mW laser power.

The temperature dependent ZFS D of the PL6 divacancy is the cornerstone of the sensing, so we measure the ODMR spectra at a magnetic field of 32.1 G. Figure 1b shows three representative ODMR spectra at different temperatures. The measured D value is 1350.6 MHz at 299.1 K, which is consistent with the ZFS of PL6 at room temperature [7,21]. Both the left and right branches of the ODMR frequencies simultaneously decrease as the temperature increases, which is due to the temperature induced decrease of ZFS D. As shown in Fig. 1c, the ZFS $D$ declines linearly as temperature rises from 280 K to 320 K. After using the linear fitting, we get the slope of $D$ versus temperature about -99.7 ± 0.4 kHz/K, which is consistent with the slope of the PL5 divacancy [10]. The slope is about 1.4 times larger than the results of NV centers in diamond [1,25,26].

After getting the temperature dependent ZFS parameter, we measure the temperature using the thermal pulse sequence including thermal Ramsey [27], thermal echo [25,26], and TCPMG [25,26] methods at a magnetic field. We first introduce the working principle of T-Rasmey. Figure 2a shows the pulse sequence of the thermal Ramsey. The subscripts -1 and +1 indicate the spin transitions between $|0\rangle \leftrightarrow |-1\rangle$ and $|0\rangle \leftrightarrow |1\rangle$, respectively. The initial state $|0\rangle$ changes to $(|0\rangle + |-1\rangle)/\sqrt{2}$ by the $\pi/2_{-1}$ pulse. Then the state evolves to $(|0\rangle + e^{-i(\varphi_D - \varphi_B)}|-1\rangle)/\sqrt{2}$ after evolution time τ, where $\varphi_D = \Delta D \cdot \tau$, $\Delta D$ is the change of the microwave detuning, $\varphi_B$ is the phase due to magnetic noise. Then the triple π pulse $\pi_{-1}\pi_{+1}\pi_{-1}$ is applied to swap the state of $|-1\rangle$ to $|1\rangle$. Finally, the state evolves to $(|0\rangle + e^{-2i\varphi_D}|1\rangle)/\sqrt{2}$ after another evolution time τ, which is the same as the NV centers in diamond [25-27]. The total phase $2\varphi_D$ is $2\Delta D \cdot \tau$, which is only related to the shifts of D, eliminating the effect of the magnetic noise [25-27]. Finally, the fluorescence signal oscillates with a frequency by $\Delta D$ [25-27]. The change of the D (temperature) can be detected by the change of the oscillation frequency. The working principle of TCPMG is similar to thermal Rasmey, except that it has more triple π pulse, which can extend the coherence time by eliminating high frequency magnetic noise [25,26].

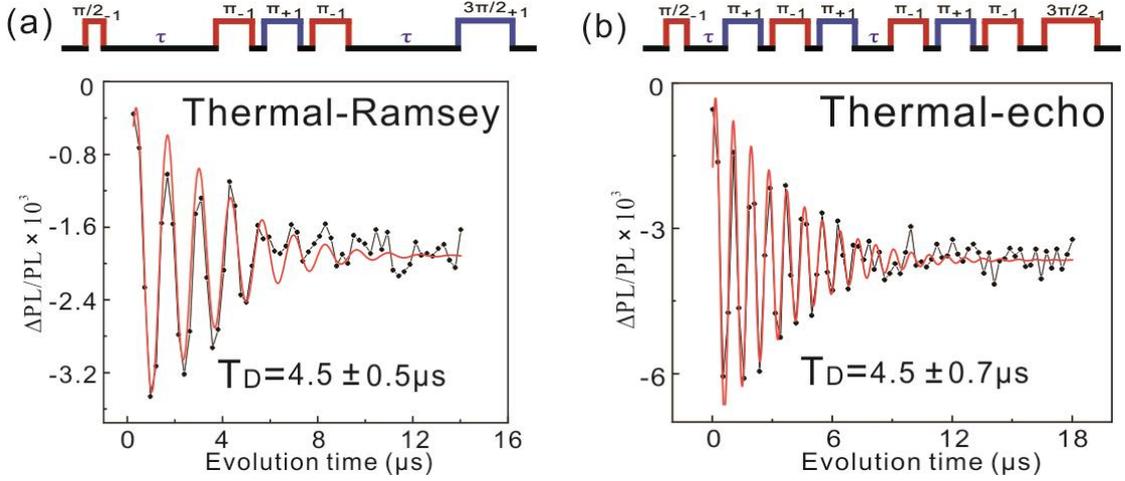

**Figure 2.** Thermal Ramsey and thermal echo at a magnetic field of 32.1 G. **(a)** Thermal Ramsey measurement. The upper part is the thermal Ramsey sequence. The red and blue lines represent the transitions of $|0\rangle \leftrightarrow |-1\rangle$ and $|0\rangle \leftrightarrow |1\rangle$, respectively. The result is shown in the lower part. **(b)** Thermal echo measurement. The upper part is the thermal echo sequence. The red lines are fittings to the data.

The experimental data of thermal Ramsey is presented at the bottom of Fig. 2a. We fit the data using a function [25-27]:

$$I = a \times exp[-(\frac{t}{T_D})^n] cos(2\pi ft + \varphi) + b \qquad (2)$$

where oscillation frequency $f = |(\omega_1 - \omega_{-1})/2 - D|$, and $\omega_1$, $\omega_{-1}$ are the microwave frequencies for transitions of $|0\rangle \leftrightarrow |1\rangle$ and $|0\rangle \leftrightarrow |-1\rangle$, respectively. $T_D$ is the coherence time of the thermal pulse. From the fitting, we get the coherence time $T_D = 4.5 \pm 0.5$ μs. We also measure the thermal echo at a magnetic field of 32.1 G, and the coherence time $T_D = 4.5 \pm 0.5$ μs, which is the same as the result of thermal Ramsey. Both the coherence times are about three times as that of $T_2^*$ [10,11], which is beneficial for temperature detection.

To further improve the coherence time for higher temperature sensing sensitivity, we perform the TCPMG methods at a magnetic of 32.1 G. Fig. 3(a) is the TCPMG-N pulse sequences in finite magnetic fields, where N is the pulse number [25,26]. Figure 3b

shows the TCPMG-1 measurement. By using the same fitting function of eq 2, we get the

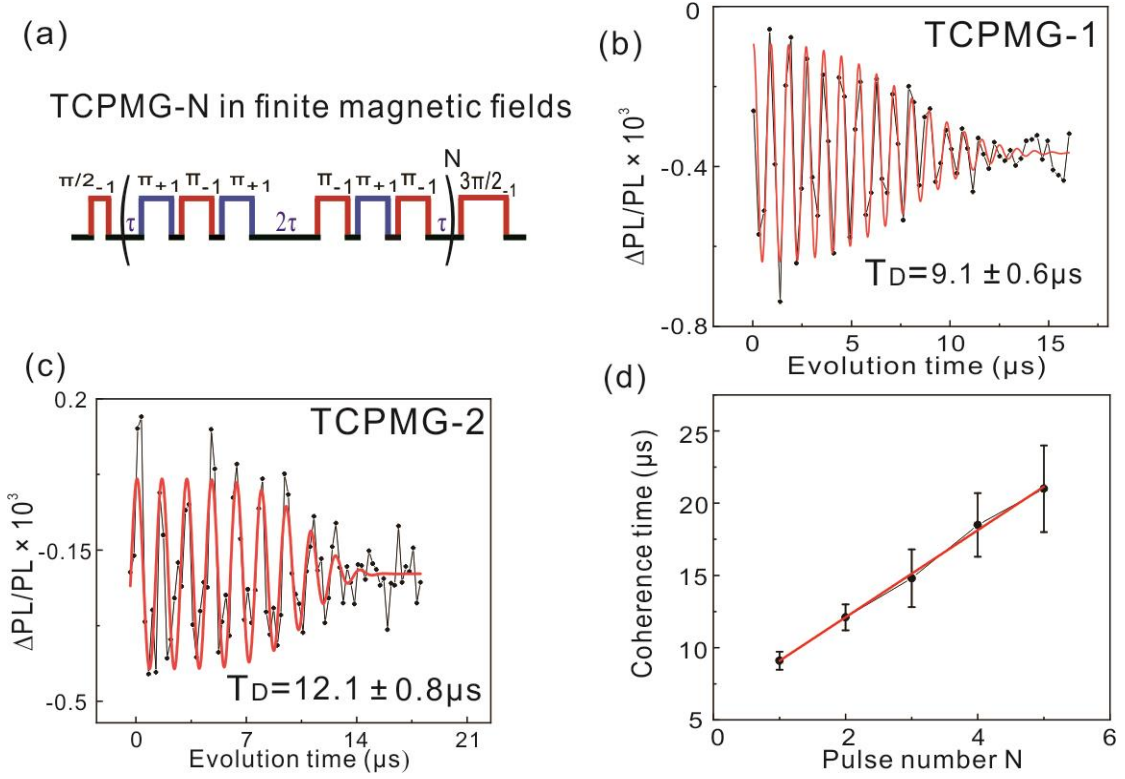

**Figure 3.** TCPMG-N pulse sequence and measurements at 32.1 G at room temperature. **(a)** The pulse sequences of the TCPMG-N. **(b)** The TCPMG-1 measurement. The coherence time is about 9.1 ± 0.6 μs. **(c)** The TCPMG-2 measurement and the corresponding coherence time increases to about 12.1 ± 0.8 μs. **(d)** The dependence of the coherence time with the pulse number N from 1 to 5. The coherence times increase linearly with the number N. The red line is a linear fitting of the data.

coherence time of 9.1 ± 0.6 μs, which is two times as the results of thermal Ramsey and thermal echo methods. The TCPMG-2 results are presented in Fig. 3c, and the coherence time is 12.1 ± 0.8 μs, which is larger than that of TCPMG-1. In Fig. 3d, we present the coherence time as a function of pulse number N. The coherence time linearly increases with N, consistent with the results of the NV centers in diamond, which demonstrates that the TCPMG methods can efficiently eliminate magnetic noise. Particularly, the coherence time of TCPMG-5 is 21 ± 3 μs, which is more than ten times

larger than the $T_2^*$. The thermometer sensitivity η can be calculated using the function:

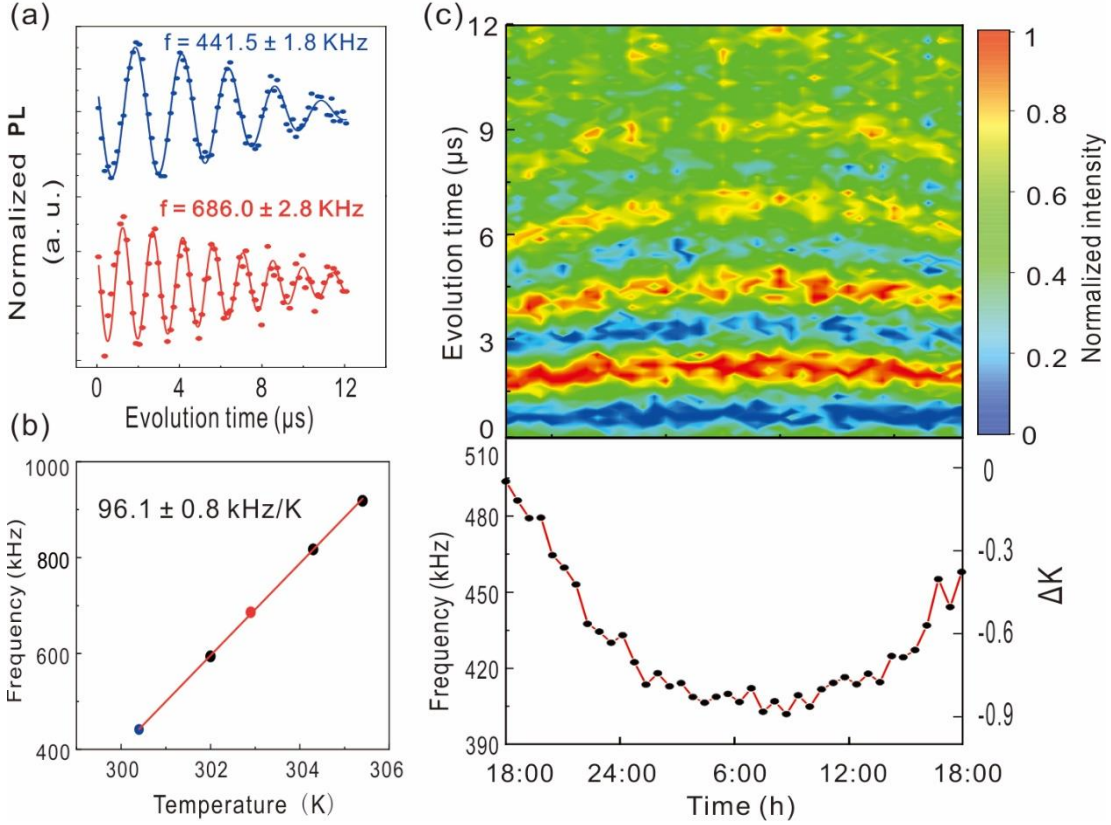

**Figure 4.** Temperature detection using TCPMG pulse sequence. **(a)** TCPMG-1 measurement results under two different heating powers. The fitting oscillation frequencies are shown in the plots. **(b)** The fitted oscillation frequencies of TCPMG-1 as a function of the temperature under five different heating powers. The red line is a linear fitting to the data. **(c)** Temperature measurements under laboratory conditions from 18:00 within 24 hours. The upper part is the continuous measurement of TCPMG-1. The lower part shows the oscillation frequency and corresponding temperature changes for 24 hours.

$$\eta = \sqrt{\frac{2(p_0+p_1)}{(p_0-p_1)^2}} \frac{1}{2\pi\frac{dD}{dT}\exp\left(-\left(\frac{t}{T_d}\right)^n\right)\sqrt{t}} \qquad (3)$$

where $p_1$ and $p_0$ are the photon counts in the dark and bright spin states per shot, respectively [10,11,25-27]. We obtain the thermometer sensitivity η of TCPMG-5 to be 13.4 mK/Hz$^{1/2}$, which is about 15 times higher than previous results [10,11]. The experiments prove the divacancy-based high sensitivity thermometer. The sensitivity can be further increased by using high order TCPMG methods, high fluorescence collection efficiency photon structure [16,28] and hybrid SiC thermometer [4].

Finally, we apply the divacancy thermometer for temperature detection using TCPMG methods with positive microwave detunings [26]. As shown in Fig. 4a, we measure the TCPMG-1 at two different temperatures, and the frequency errors are about 4 times less than previous Ramsey methods [10,11], indicating high precision temperature sensing of TCPMG methods [26,27]. Moreover, the oscillation frequency is larger at higher temperature due to the decrease of D with increased temperature. The oscillation frequency with respect to the temperature is presented in Fig. 4b. The oscillation frequency linearly increases with temperature at a slope of -96.1 ± 0.8 kHz/K, which agrees with previous ODMR results. Then we utilize the thermometer to track the changes of laboratory temperature over 24 hours from 18:00, and the results are presented in Fig. 4c. The upper part shows the continuous TCPMG-1 signals, and the oscillation frequencies change with time. The lower part shows the oscillation frequency and relative temperature changes for 24 hours. We can conclude that the laboratory temperature drops slowly at night and keeps stable in the daytime, then rises after 16:00, which is in line with outdoor temperature.

In conclusion, we realize a high sensitivity SiC PL6 divacancy-based thermometer using TCPMG methods. The temperature dependent ODMR measurements show that the ZFS D of PL6 is about -99.7 ± 0.4 kHz/K. On this basis, we perform the thermal Ramsey, thermal echo, and TCPMG methods under a magnetic field. The coherence time of the TCPMG linearly increases with pulse number N, and the coherence time of TCPMG-5 is 21 ± 3 μs, which is more than ten times larger than the $T_2^*$. The corresponding thermometer sensitivity is 13.4 mK/Hz$^{1/2}$, which is about 15 times as the previous Ramsey methods[10,11]. Finally, we apply the thermometer to trace the laboratory temperature for one day, which is consistent with outdoor temperature. The experiments pave the way for the high sensitivity SiC divacancy-based thermometer applications in various areas of modern technology and science.

**Notes**

The authors declare no competing financial interest.

**Acknowledge**

This work is supported by the National Natural Science Foundation of China (Grants 61905233, 11975221). J. F. Wang also acknowledges financial support from the Science Specialty Program of Sichuan University (Grand No. 2020SCUNL210).